\documentclass{jfm}
\usepackage{graphicx}
\usepackage{epstopdf, epsfig, amsmath}
\usepackage{bm}

\DeclareMathOperator{\sech}{sech}

\shorttitle{Coherent structure coloring}
\shortauthor{K.L. Schlueter-Kuck and J. O. Dabiri}

\title{Model parameter estimation using coherent structure coloring}

\author{Kristy L. Schlueter-Kuck\aff{1}
  \and John O. Dabiri\aff{1,2}
\corresp{\email{jodabiri@stanford.edu}}}

\affiliation{\aff{1}{Department of Mechanical Engineering, Stanford University, Stanford, CA 94305, USA}
\aff{2}Department of Civil and Environmental Engineering, Stanford University, Stanford, CA 94305, USA}

\begin{document}

\maketitle

\begin{abstract}

Lagrangian data assimilation is a complex problem in oceanic and atmospheric modeling. Tracking drifters in large-scale geophysical flows can involve uncertainty in drifter location, complex inertial effects, and other factors which make comparing them to simulated Lagrangian trajectories from numerical models extremely challenging. Temporal and spatial discretization, factors necessary in modeling large scale flows, also contribute to separation between real and simulated drifter trajectories. The chaotic advection inherent in these turbulent flows tends to separate even closely spaced tracer particles, making error metrics based solely on drifter displacements unsuitable for estimating model parameters. We propose to instead use error in the coherent structure coloring (CSC) field to assess model skill. The CSC field provides a spatial representation of the underlying coherent patterns in the flow, and we show that it is a more robust metric for assessing model accuracy. Through the use of two test cases, one considering spatial uncertainty in particle initialization, and one examining the influence of stochastic error along a trajectory and temporal discretization, we show that error in the coherent structure coloring field can be used to accurately determine single or multiple simultaneously unknown model parameters, whereas a conventional error metric based on error in drifter displacement fails. Because the CSC field enhances the difference in error between correct and incorrect model parameters, error minima in model parameter sweeps become more distinct. The effectiveness and robustness of this method for single and multi-parameter estimation in analytical flows suggests that Lagrangian data assimilation for real oceanic and atmospheric models would benefit from a similar approach.
\end{abstract}

\section{\label{sec:intro}Introduction}

Models of oceanic and atmospheric geophysical flows are of great importance in weather prediction and nowcasting~\citep{Kalnay2002}, understanding the effects of natural disasters such as tsunamis~\citep{Ioualalen2007, Yamazaki2011} and anthropogenic disasters such as oil spills~\citep{Mariano2011}, and quantifying transport of nutrients and passive tracers such as heat and salt in the ocean~\citep{Boning1994}.  Due to the large length scales of such flows, they are frequently measured directly by sparse sensors that approximately follow the flow, including ocean drifters and weather balloons.  These sensors provide valuable data on flow velocities, temperature, density, and other parameters that can be used to validate and tune the numerous parameters in the complex models developed to simulate these flows.  However, the integration of the real Lagrangian flow data into typical Eulerian model descriptions is complicated by the nonlinear relationship between the underlying flow velocities and the path taken by individual parcels of fluid or fluid tracers.  In flows with significant chaotic advection, even small uncertainties in spatial or temporal initial conditions can lead to large divergence in the trajectories taken by tracers that are initially closely spaced~\citep{Guckenheimer1983}.  Further, real drifters experience effects such as added mass, Basset forces, and windage that influence their trajectories~\citep{Putman2013, Olivieri2014}.  Because oceanic and atmospheric models frequently use error in physical positions of simulated Lagrangian particles compared to real drifters in the flow to assess the accuracy of the models~\citep{Apte2008}, the combination of these factors can complicate parameter determination leading to large errors even for physically correct models.  All of the factors noted previously need to be accounted for when comparing real drifters to simulated drifters for the purposes of Lagranginan data assimilation.  Often times, even a small uncertainty in these factors can lead to large departures of the simulated trajectories from the real drifters.  It is therefore useful to make sure any method for trajectory comparison is robust to reasonable levels of uncertainty.

This issue necessitates the development of error metrics that are robust to the effects of chaotic advection and uncertainties in drifter position.  Recent advances in coherent pattern identification provide a potential solution.  One goal of coherent pattern (or coherent structure) identification is to gain insight into transport barriers in flows.  This can be accomplished by identifying lines or surfaces of maximal separation of initially closely spaced particles, as in finite time Lyapunov exponent (FTLE) analysis~\citep{Haller2000, Shadden2005}, by clustering trajectories using machine learning algorithms~\citep{Froyland2015}, identifying groups of trajectories that remain spatially compact~\citep{Hadjighasem2016}, bounding intertwining lines in a hybrid spatial/temporal parameter space\citep{Thiffeault2010}, or clustering trajectories based on their relative dissimilarities~\citep{Schlueter-Kuck2017a}.  With the exclusion of FTLE analysis, these techniques have been developed to be compatible with sparsely initialized tracer distributions, making them well suited for analysis of ocean drifters and weather balloons.  The underlying flow structure represented by these coherent patterns can potentially provide an effective metric for quantifying error in models.

Recently, one innovative study combined these ideas of coherent structure identification and model parameter estimation for Lagrangian data assimilation.  In this study, the authors used principal component analysis (PCA) of the evolution of the horizontal position of a set of drifters initialized around the center of two gyre cores in an analytical flow field~\citep{Maclean2017}.  They successfully identified model parameters when a random component was added to the advected drifter location at every time step.  However, this study did not address several factors that complicate the use of GPS enabled drifters to map ocean currents.  Namely, while the positions of drifters may be known with high accuracy, even minimal error in the initial location of the drifters can significantly impact the drifter trajectory.  Furthermore, due to the difficulty in deploying drifters simultaneously, irregular or random initialization of drifters must be accounted for.  Here, we propose a method of parameter estimation based on coherent structure coloring, which uses the dissimilarity in the drifter trajectories to identify the underlying coherent flow patterns.  This method has been shown to be robust to measurement uncertainty and data loss, is effective even in instances of sparse data (approximately 300 drifters tracked in a two-dimensional flow), and is effective at identifying coherent sets of trajectories~\citep{Schlueter-Kuck2017b}.


The analysis here is focused on two specific test cases.  In the first test case, the effects of error in spatial initialization are examined.  In the second case, effects on parameter estimation due to temporal discretization, error in spatial initialization, and stochastic error along the drifter trajectories (meant to represent any number of inertial effects and/or spatial discretization), are quantified.  Section 2 details the computational setup of the analysis, and section 3 presents results of the analysis.  Conclusions and future directions are discussed in section 4.

\section{Methods}
\subsection{Test Case 1: Drifter Position Uncertainty}
The first test case examined here seeks to quantify the effect of uncertainty in the initial location of Lagrangian drifters on the parameter estimation process.  Because chaotic advection in fluid flows tends to separate even closely spaced fluid particles over sufficiently long periods of time, even a small uncertainty in spatial initialization can lead to large particle displacement errors (e.g. ~\citet{Putman2013}).  For this case, the analytical quadruple gyre, governed by equations~\ref{eqn:quadgyre1}--~\ref{eqn:quadgyre3}, was used to model the underlying flow:
\begin{eqnarray}\label{eqn:quadgyre}
u_x = \frac{dx}{dt} & =&-\pi \alpha\sin(\pi f)\cos(\pi y)\label{eqn:quadgyre1} \\
u_y = \frac{dy}{dt} & =&\pi \alpha\cos(\pi f)\sin(\pi y)(2ax+b) \label{eqn:quadgyre2}
\end{eqnarray}
where $x$ and $y$ are the spatial coordinates, $t$ is time, and
\begin{eqnarray}\label{eqn:quadgyre3}
a =\epsilon\sin(\omega t), 
b =1-2\epsilon\sin(\omega t), 
f =ax^2+bx.
\end{eqnarray}
Parameter values of $\alpha=0.1$, $\epsilon=0.1$, and $\omega=2\pi/10$ were used, giving the flow a periodic east-west oscillation.

To build a set of ``real" trajectories, analogous to an array of drifter positions at discrete moments in time in an oceanic or atmospheric flow, a set of 500 drifters was randomly initialized in the domain $x=[0,2]$, $y=[-1,1]$.  These drifters were advected using a fifth-order Runge-Kutta integration scheme with a relative error tolerance of $10^{-6}$ and an absolute error tolerance of $10^{-9}$.  These parameters were used for calculation of all ``real" trajectories in this study.  The ``simulated'' particles were initialized near the initial locations of the real trajectories, with an offset of $0.18$ in a random direction, corresponding to $9\%$ of the length of the domain.  For the single-parameter analysis, the parameter $\epsilon$ in equation~\ref{eqn:quadgyre3} was assumed to be unknown, and a set of simulated trajectories was created for 201 equally-spaced values of $\epsilon$ in the range $\epsilon=[0,0.4]$.  For each value of $\epsilon$, ten independent simulations with different initial conditions were run, in order to examine the repeatability of the results.  For the multiple-parameter analysis, $\alpha$, $\omega$, and $\epsilon$ in equations~\ref{eqn:quadgyre1}--\ref{eqn:quadgyre3} were assumed to be unknown, and a set of simulated trajectories was created for all combinations of 21 equally-spaced values of $\alpha$ in the range $\alpha=[0,0.4]$, 21 equally-spaced values of $\omega$ in the range $\omega=[0,\pi]$, and 21 equally-spaced values of $\epsilon$ in the range $\epsilon=[0,0.4]$.  Each unique combination of parameters was tested using ten independent simulations, resulting in 92,610 simulations of 500 particles each for the multi-parameter study.

\subsection{Test Case 2: Temporal Discretization in Model and Stochastic Position Error Along Trajectories}
The second test case in this study examined the combined effects of model temporal discretization and stochastic error on parameter estimation.  Because there are many factors that influence the divergence of simulated drifter trajectories in models from the real, observed drifter trajectories, it is critical that any parameter estimation scheme be able to account for the complex interaction of several of these factors.  This test utilized the analytical Bickley jet flow, frequently used as a model of zonal atmospheric currents~\citep{Rypina2007}, which is defined by the stream function $\psi=\psi_0+\psi_1$, where
\begin{eqnarray}
\psi_0 & =&c_3y-UL\tanh\left(y/L\right)\label{eqn:bickley_eqn1a}\\
\psi_1 & =&UL\sech^2\left(y/L\right)\sum_{n=1}^3\epsilon_n\cos\left(k_n\left(x-\sigma_nt\right)\right)\label{eqn:bickley_eqn1b}
\end{eqnarray}

For this analysis, $U=62.66$ ms$^{-1}$, $L=1770$ km, $k_n=2n/r_0$, $c=[0.1446U$, $0.205U$, $0.461U]$,  $\sigma=c-c(3)$, and $\epsilon=[0.0075$, $0.15$, $0.3]$, and the flow is computed on the interval $x=[0$, $20\times10^6]$ m, $y=[-3\times10^6$, $3\times10^6]$ m, over the time interval $t=[0$, $80]$ days.

For this test, the ``real'' trajectories were calculated by initializing 500 particles in the domain and advecting them using a fifth-order Runge-Kutta scheme, with the same error tolerances as noted previously.  The drifters were advected over 80 days according to equations~\ref{eqn:bickley_eqn1a} and~\ref{eqn:bickley_eqn1b}.

For this analysis, the ``simulated'' particles were initialized near the initial locations of the real drifters, offset by 60 km (0.3\% of the horizontal spatial domain) in a random direction.  Although 60 km is more uncertainty than could be expected from GPS data alone, this value is meant to capture some of the effect of spatial discretization in the model as well, which was not considered directly.  A random drifter initialized in the Bickley jet flow at $t=0$ with a spatial resolution of 100 km will on average, accumulate 60 km of error when compared to a ``real" drifter after one 30 minute time step if linear interpolation between grid points is used to estimate velocity.  The particles were advected using Euler integration with a temporal discretization of 30 minutes.  At every time step, a stochastic error with a standard deviation of 60 meters was added to the drifter position.  For this analysis, the parameter $\epsilon(3)$ was assumed to be unknown, and a set of simulated trajectories was created for 201 equally-spaced values of $\epsilon(3)$ in the range $\epsilon(3)=[0,1]$.

\subsection{Coherent Structure Coloring and Error Quantification}
In each test case above, the underlying coherent flow pattern was quantified using the CSC vector calculated for the ``real'' trajectories and each set of simulated trajectories~\citep{Schlueter-Kuck2017a}.  In the CSC algorithm, dissimilarity between two particle trajectories is represented numerically using a weighted adjacency matrix $A$, where $a_{ij}$ contains the weight of the edge connecting particle $i$ and particle $j$:
\begin{equation}
a_{ij}=\frac{1}{\overline{r_{ij}}T^{1/2}}\left[\sum_{k=0}^{T-1}(\overline{r_{ij}}-r_{ij}(t_k))^2\right]^{1/2}
\end{equation}
where $r_{ij}(t_k)$ is the distance between two particles $i$ and $j$ at time $t_k$, and $\overline{r_{ij}}$ is the average distance between the two fluid particle trajectories.  Conceptually, $a_{ij}$ quantifies the standard deviation of the distance between particle trajectories normalized by their average spacing.  The corresponding generalized eigenvalue problem that quantifies the difference between dissimilar particles is 
\begin{equation}
LX=\lambda DX
\end{equation}
where 
\begin{equation}
d_{ij}= \left\{
\begin{array}{ll}
    0,			& i\neq j\\
    \sum_{k=1}^N a_{ik},       & i=j,
\end{array} \right.
\label{eqn:eigenvalue}
\end{equation}
and $L=D-A$ is the graph Laplacian.  In order to maximize the differences between dissimilar particles, $X_1=X$ is the eigenvector associated with the maximum eigenvalue, $\lambda_1$, of this problem, under the constraint that $X^\prime DX$ remains finite.  Each element of $X_1$ assigns that value of CSC to the corresponding fluid particle.  This vector can be visualized in the spatial domain by mapping each element of the CSC vector to the initial particle location, and interpolating between particles to obtain a contour field.

Two error metrics were examined for each test case.  The first metric, the average particle displacement error, is defined by
\begin{equation}\label{eqn:disp_error}
E_{disp}=\frac{1}{NT}\sum_{i=1}^N\sum_{j=1}^T|\bm{x}_{real}^{i,j}-\bm{x}_{sim}^{i,j}|,
\end{equation}
where $N$ is the number of drifters tracked, $T$ is the number of time steps, and $\bm{x}_{real}^{i,j}$ and $\bm{x}_{sim}^{i,j}$ are the spatial location of the $i^{th}$ real drifter and simulated drifter, respectively, at the $j^{th}$ time step.

The error in the CSC field is calculated as follows:
\begin{equation}\label{eqn:CSC_field_error}
E_{CSC_f}=\frac{1}{M}\sum_{i=1}^M\left|\mathrm{CSC}_{f,real}^i-\mathrm{CSC}_{f,sim}^i\right|,
\end{equation}
where $\mathrm{CSC}_{f, real}^i$ and $\mathrm{CSC}_{f, sim}^i$ are the CSC field values of the $i^{th}$ Cartesian grid node for the set of real and simulated sets of drifters, respectively, and $M$ is the number of Cartesian grid nodes in the CSC field.  It is important to note that the error in the interpolated CSC field was found to be more effective for parameter estimation than the CSC vector itself, as it is more robust to individual particle position.  This point will be expanded upon in the results section.  The CSC field was calculated by interpolating the CSC vector onto a grid of spatial locations using triangulation-based linear interpolation, and $M$ is the product of the number of discrete horizontal locations and the number of discrete vertical locations.  The error based on the CSC field was found to be insensitive to the interpolation grid coarseness, as long as the average spacing between grid elements was smaller than the average spacing between particles tracked.  The CSC vector error, shown in the results section for comparison, is quantified as follows:
\begin{equation}\label{eqn:CSC_vector_error}
E_{CSC_v}=\frac{1}{N}\sum_{i=1}^N\left|\mathrm{CSC}_{v, real}^i-\mathrm{CSC}_{v, sim}^i\right|,
\end{equation}
where $\mathrm{CSC}_{v, real}^i$ and $\mathrm{CSC}_{v, sim}^i$ are the CSC vector values of the $i^{th}$ drifter for the set of real and simulated sets of drifters, respectively, and $N$ is the total number of drifters tracked.

\section{Results}
\subsection{Test Case 1: Drifter Position Uncertainty}
\begin{figure}
\centering
\includegraphics[width=\linewidth]{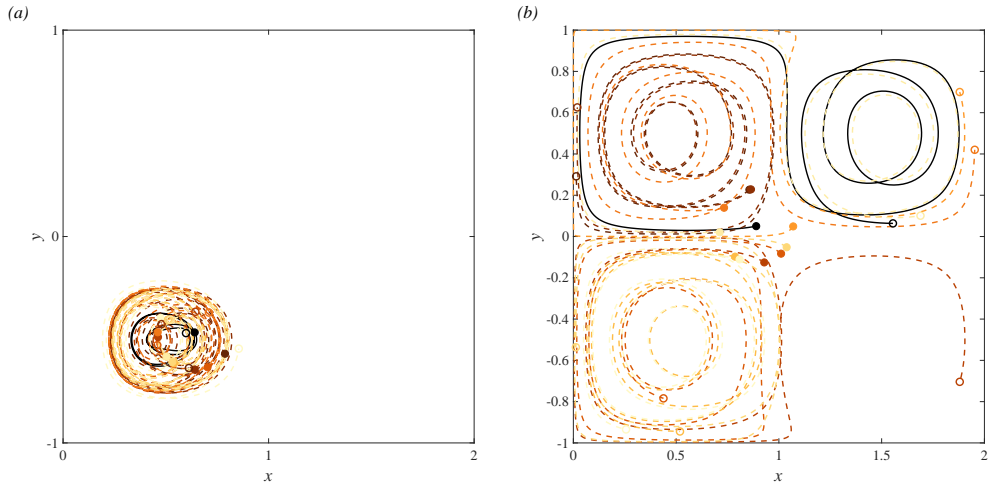}
\caption{Sample drifters in quadruple gyre flow.  Solid black lines indicate ``real'' drifter trajectories and dashed lines of other colors indicate simulated drifter trajectories.  Closed circles correspond to drifter initial locations at $t_0=2.5$ and open circles to drifter final locations at $t_f=42.5$. (a) Drifter initialized in the center of the lower left coherent vortex. (b) Drifter initialized near the intersection of the four quadrants.}
\label{fig:quadgyre1}
\end{figure}

The first test case examined the effect of error in initial drifter position on the ability to accurately determine the parameter $\epsilon$, the magnitude of the periodic horizontal oscillation in the quadruple gyre flow.  Figure~\ref{fig:quadgyre1} examines the ``real" and simulated trajectories (with the correct value of the unknown parameter, $\epsilon=0.1$ but slightly errant initial position) for two initial positions, one within the lower left coherent vortex, and the second near the transport barrier located at the intersection of the four quadrants.  It is clear that by the end of the simulated time interval of four horizontal oscillation cycles, chaotic advection in the flow has strongly separated the real drifter located near the center of the domain (i.e. figure~\ref{fig:quadgyre1}(b), black trajectory) from most of the simulated trajectories of the same drifter.  In fact, there are simulated trajectories that end up in each of the four quadrants. This separation is due to the error in initial position resulting in the particles being initialized in a different quadrant of the flow.  Subsequently, the transport barrier separating the ``real" particle from the corresponding simulated particles leads to an exponentially large spatial separation in time.  In contrast, for the drifter initialized at the center of the lower left vortex, the simulated trajectories remain compact, as there are no transport barriers separating the ``real" drifter initial position from the simulated initial positions.

\begin{figure}
\centering
\includegraphics[width=\linewidth]{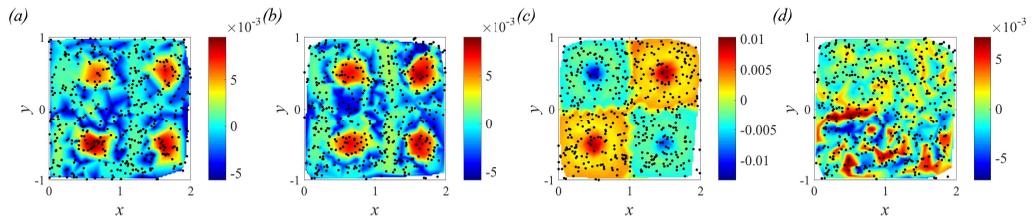}
\caption{CSC contours for quadruple gyre overlaid with dots indicating initial positions of 500 drifters for $t=[2.5$, $42.5]$. (a) ``Real'' trajectories. (b) Simulated trajectories with errant initial position but correct value of $\epsilon=0.1$. (c) Simulated trajectories with errant initial position and with incorrect value of $\epsilon=0$. (d) Simulated trajectories with errant initial position and with incorrect value of $\epsilon=0.4$.}
\label{fig:quadgyre2}
\end{figure}

Figure~\ref{fig:quadgyre2} shows the CSC fields for the real drifter trajectories (a), the simulated trajectories with the correct value of epsilon (b), and the simulated trajectories with incorrect values of $\epsilon=0$ (c) and $\epsilon=0.4$ (d).  For the simulated case with the correct value of epsilon, the CSC field highlights the largest kinematic dissimilarity in the flow, between the particles that remain in the gyre cores (with high CSC values) and those particles that switch quadrants during the prescribed time interval (with low CSC values).  By comparing the CSC fields for the real and correctly simulated sets, it is evident that the CSC field is robust to chaotic advection of individual drifters.  This is because the Lagrangian drifters are used as landmarks for interpolating the underlying coherent patterns in the flow, but do not individually dictate these patterns.  It should be noted that the magnitude of CSC vector and field values is dependent on the number of particles used for analysis.  Thus, when assessing relative error values, the number of particles tracked must be held constant.  Furthermore, because the CSC vector is an eigenvector associated with the generalized eigenvalue problem given by ~\ref{eqn:eigenvalue}, both the calculated CSC vector and its negative will need to be considered when comparing the real and simulated drifter sets, and the minimum error kept.

Because of the factors noted above, two independently initialized sets of particles \textit{should} have the same CSC field, up to error on the order of particle separation distance, regardless of how different the spatial initializations are.  The CSC field for the incorrect value of $\epsilon=0$ clearly has a different underlying flow structure, where the largest kinematic dissimilarity is between the quadrants that have counterclockwise rotation and those with clockwise rotation of fluid.  Similarly, the CSC field for the incorrect value of $\epsilon=0.4$ also has a flow structure distinct from the ``real" flow; the magnitude of the left-right oscillation is strong enough to eliminate the coherent vortices in each quadrant, and the flow mixes chaotically.  This example highlights why using the error in the interpolated CSC field is more robust than using the error in the CSC vector.  A drifter just outside the boundary of a coherent structure will be assigned a different CSC value than a drifter just inside the boundary, but the location of the boundary itself in the CSC field will be the same regardless of the location at which that particular drifter is initialized.  By moving away from an individual trajectory analysis, and toward a field-based analysis of the underlying flow, correct model parameters can be identified unambiguously.

\begin{figure}
\centering
\includegraphics[width=\linewidth]{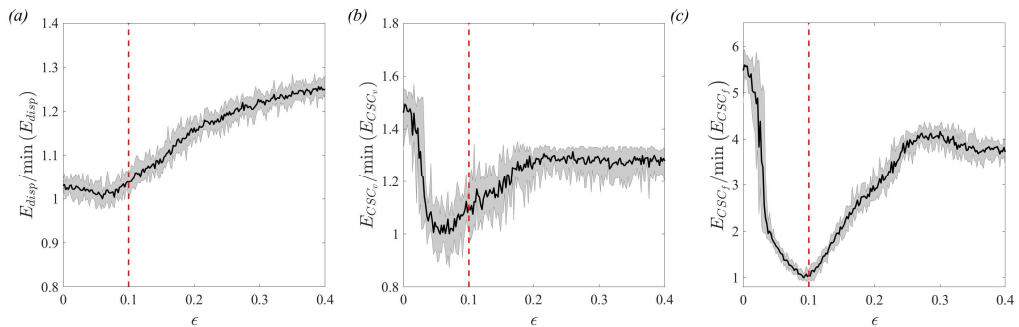}
\caption{Normalized error versus candidate parameter $\epsilon$.  Error metrics are normalized by the lowest error over the range of $\epsilon$ tested.  Solid black line indicates the error averaged over 10 individual sweeps, and shaded region bounds the range of errors seen for an individual sweep.  Red dotted line indicates the correct value for the parameter $\epsilon$. (a) Particle displacement error, i.e. eq.~\ref{eqn:disp_error}. (b) CSC vector error, i.e. eq.~\ref{eqn:CSC_vector_error}. (c) CSC field error, i.e. eq.~\ref{eqn:CSC_field_error}.}
\label{fig:quadgyre3}
\end{figure}

The conventional particle displacement error and the CSC field error are compared in figure~\ref{fig:quadgyre3} as a function of the selected value of $\epsilon$ for the simulated particle set.  While the CSC error is minimized at the correct value of $\epsilon=0.1$, the particle displacement error and CSC vector error are not.  Hence, model parameter estimation using the CSC field recovers the correct parameter value despite the potentially confounding effects of initial position uncertainty.  The conventional error metric fails under the same conditions.  It is also interesting to note that the CSC field error has a larger difference in error between the correct and incorrect states, with the error at $\epsilon=0$ exceeding five times the error for $\epsilon=0.1$.  In contrast, the particle displacement error varies less than $30\%$ from the minimum error over the range of parameters tested.  This means that not only does the CSC field aid in accurately predicting the correct model parameter, but it amplifies the error between correct and incorrect parameter values.

The purpose of the multi-parameter analysis was to evaluate the effectiveness of this method for determining more than one unknown parameter simultaneously, a problem that is frequently encountered in oceanic and atmospheric models.  An analysis of the full parameter space, while not computationally efficient, was used here for completeness and visualization purposes.  It is beyond the scope of this work to investigate various optimization techniques for determining the global error minimum, although any viable techniques would need to take into account the noisiness of the error signal as well as its non-convexity.  These factors make common approaches such as gradient descent optimization unfeasible.

\begin{figure}
\centering
\includegraphics[width=\linewidth]{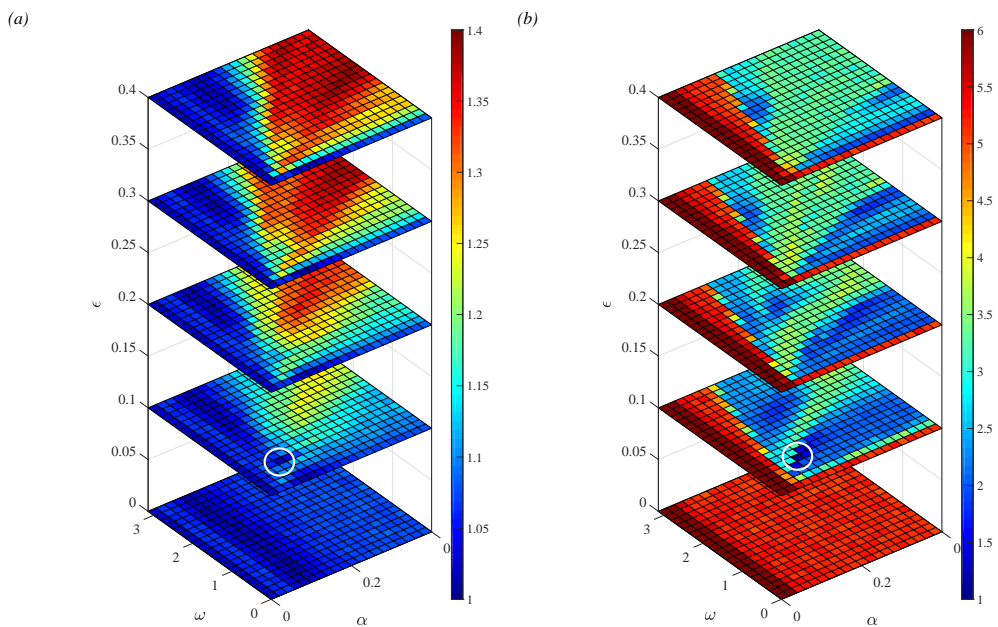}
\caption{Normalized error metrics averaged over ten individual three-parameter sweeps in $\alpha$, $\omega$, and $\epsilon$.  Error metrics are normalized by the lowest error over the range of all three parameters tested.  (a) Displacement error for selected $\epsilon$-slices.  (b) CSC field error for selected  $\epsilon$-slices.  White circles indicate the approximate location of the global minimum.}
\label{fig:multiparam1}
\end{figure}

\begin{figure}
\centering
\includegraphics[width=\linewidth]{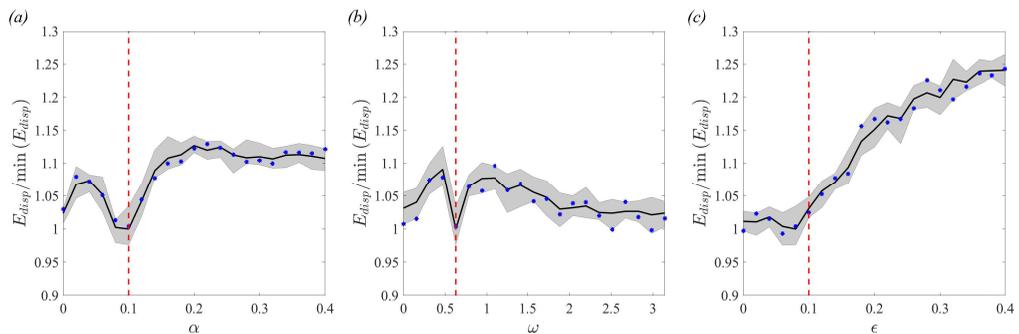}
\caption{Normalized displacement error slices through the three-parameter sweep.  Solid black line indicates the error averaged over 10 individual sweeps, and shaded region bounds the range of errors seen for an individual sweep. Blue stars indicate error for a single representative sweep.  Dotted red lines indicate the correct value of the parameter.  (a) Slice at $\omega=2\pi/10$, $\epsilon=0.08$. (b) Slice at $\alpha=0.10$, $\epsilon=0.08$. (c) Slice at $\alpha=0.10$, $\omega=2\pi/10$.}
\label{fig:multiparam2}
\end{figure}

\begin{figure}
\centering
\includegraphics[width=\linewidth]{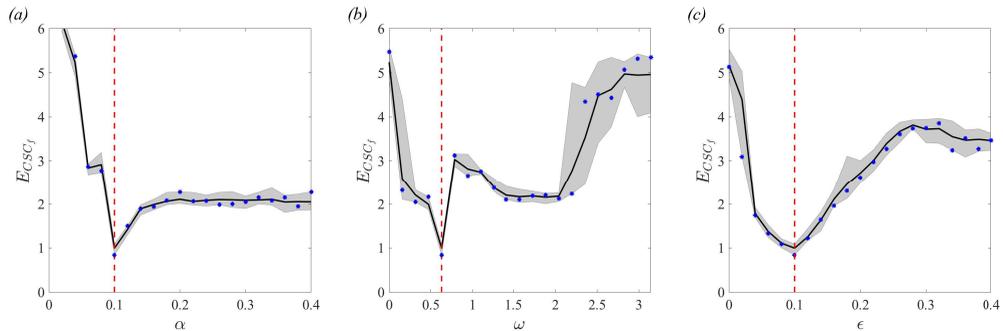}
\caption{CSC field error slices through the three-parameter sweep.  Solid black line indicates the error averaged over 10 individual sweeps, and shaded region bounds the range of errors seen for an individual sweep. Blue stars indicate error for a single representative sweep.  Dotted red lines indicate the correct value of the parameter.  (a) Slice at $\omega=2\pi/10$, $\epsilon=0.10$. (b) Slice at $\alpha=0.10$, $\epsilon=0.10$. (c) Slice at $\alpha=0.10$, $\omega=2\pi/10$.}
\label{fig:multiparam3}
\end{figure}

Figure~\ref{fig:multiparam1} shows the average particle displacement error (a) and average CSC field error (b) as a function of the three unknown parameters, $\alpha$, $\omega$, and $\epsilon$, averaging the error produced for each unique combination of parameters over ten independent simulations with different initial offsets for each drifter.  Data for only five of the 21 values of $\epsilon$ is displayed for simplicity.  The location of the minimum in the error is highlighted by the white circles for each case.  It is clear that both the particle displacement error and the CSC field error exhibit global minima in approximately the same region of the parameter space.  
The error minima are examined more closely in table~\ref{table:multiparam_table1}, which shows the parameter values where the displacement, CSC vector, and CSC field error are minimized for each of the individual ten parameter sweeps as well as for the averaged error field for each metric.  These data show that while averaging data from ten sweeps results in a nearly-correct identification of all three unknown parameters for both the displacement error and the CSC field error, individual sweeps of the displacement error result in wildly incorrect parameter identification, while the CSC field error resulted in a global minimum at the  correct parameter-space location for every single individual parameter sweep.  The tenfold-reduction in the number of simulations needed for correct parameter identification using the CSC field error is potentially extremely beneficial for complex and computationally expensive simulations of oceanic and atmospheric flows, where large simulation ensembles are currently required in practice~\citep{Kalnay2002}.
\begin{table}[]
\begin{tabular}{p{0.5 in}p{0.5 in}p{1.2 in}p{1.2 in}p{1.2 in}}
     & error metric & \multicolumn{1}{c}{$E_{disp}$} & \multicolumn{1}{c}{$E_{CSC_v}$} & \multicolumn{1}{c}{$E_{CSC_f}$} \\
run \#  &              &            &                         &                                      \\\hline
average &         &     (0.10, $2\pi/10$, 0.08)       &      (0.18, $5\pi/20$, 0.16)                   &   (0.10, $2\pi/10$, 0.10) \\\hline

1       &              &     (0.08, $2\pi/10$, 0.08)       &      (0.18, $5\pi/20$, 0.08)                   &   (0.10, $2\pi/10$, 0.10) \\
2       &              &     (0.10,         0     , 0.10)       &      (0.38 $7\pi/20$, 0.30)                    &   (0.10, $2\pi/10$, 0.10) \\
3       &              &     (0.10, $2\pi/10$, 0.08)       &      (0.28, $5\pi/20$, 0.24)                   &   (0.10, $2\pi/10$, 0.10) \\
4       &              &     (0.06, $7\pi/20$, 0.04)       &      (0.26, $5\pi/20$, 0.28)                   &  (0.10, $2\pi/10$, 0.10) \\
5       &              &     (0.08,          0    , 0.34)       &      (0.22, $5\pi/20$, 0.18)                   &   (0.10, $2\pi/10$, 0.10) \\
6       &              &     (0.08, $19\pi/20$, 0.12)     &      (0.26, $3\pi/10$, 0.24)                   &   (0.10, $2\pi/10$, 0.10) \\
7       &              &     (0.10, $2\pi/10$, 0.08)       &      (0.20, $5\pi/20$, 0.14)                   &   (0.10, $2\pi/10$, 0.10) \\
8       &              &     (0.08, $\pi       $, 0.26)       &      (0.18, $2\pi/10$, 0.20)                   &   (0.10, $2\pi/10$, 0.10) \\
9       &              &     (0.10, $2\pi/10$, 0.08)       &      (0.36, $7\pi/20$, 0.22)                   &   (0.10, $2\pi/10$, 0.10) \\
10      &             &     (0.10, $2\pi/10$, 0.06)       &      (0.36, $3\pi/10$, 0.24)                   &   (0.10, $2\pi/10$, 0.10)
\end{tabular}
\label{table:multiparam_table1}
\caption{Global minima, ($\alpha_{min}$, $\omega_{min}$, $\epsilon_{min}$), for each of ten individual three-parameter sweeps, along with the global minimum in the error averaged over all ten sweeps.  }
\end{table}
Figure~\ref{fig:multiparam2} shows slices of the error field for the displacement error at the parameter values where the global minimum was identified, in this case at  ($\alpha_{min}$, $\omega_{min}$, $\epsilon_{min}$)=(0.10, $2\pi/10$, 0.08).  Data from one of the individual sweeps is plotted along with the average of all 10 sweeps and the range of error values resulting for each parameter combination.
Figure~\ref{fig:multiparam3} shows comparable data for the CSC field error, with slices taken at ($\alpha_{min}$, $\omega_{min}$, $\epsilon_{min}$)=(0.10, $2\pi/10$, 0.10).  It is clear from these plots that there is less variation among the ten sweeps for the CSC field error, especially in regions of the parameter space close to the global minimum.  This feature of the CSC field error allows for correct identification of all of the unknown parameters simultaneously, without the need for an ensemble of multiple simulation iterations.

\subsection{Test Case 2: Temporal Discretization in Model and Stochastic Position Error Along Trajectories}
\begin{figure}
\centering
\includegraphics[width=\linewidth]{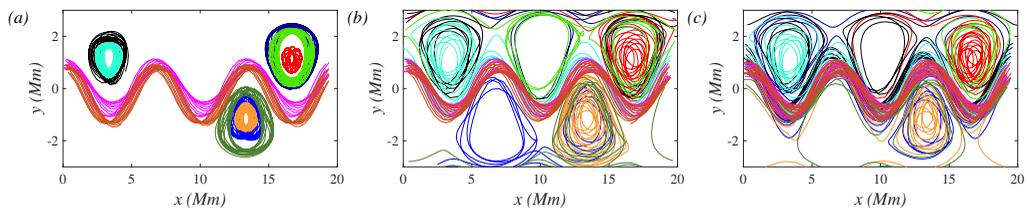}
\caption{Ten drifters in the Bickley jet flow. (a) ``Real'' trajectories. (b) Trajectories with temporal discretization, no stochastic position error, and correct value of $\epsilon(3)=0.3$. (c) Trajectories with temporal discretization, stochastic position error, and correct value of $\epsilon(3)=0.3$}.
\label{fig:bickley1}
\end{figure}

\begin{figure}
\centering
\includegraphics[width=\linewidth]{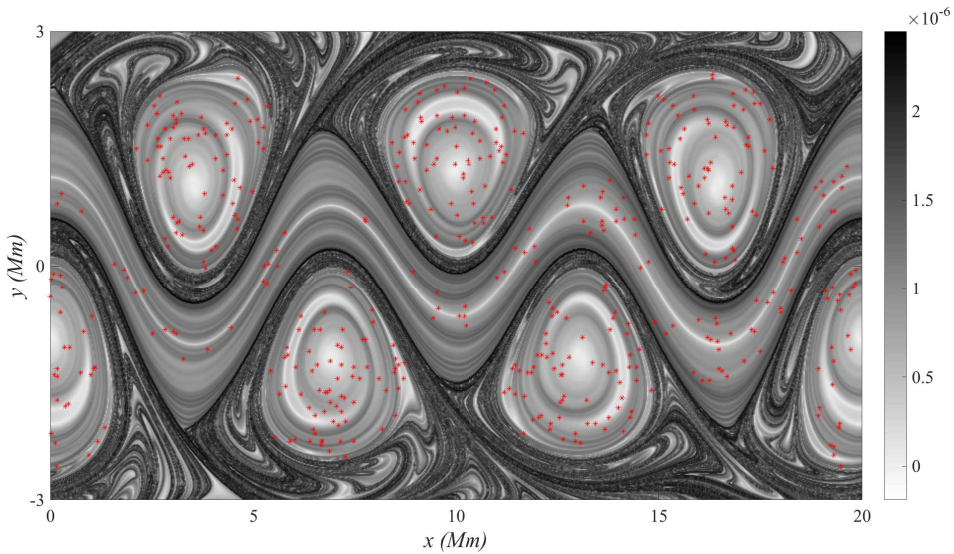}
\caption{Initial positions for the ``real" drifters in the Bickley jet study, indicated by the red stars.  Background shows the FTLE field calculated over the time interval t=[0, $3456\times10^3$] s.}
\label{fig:bickley2}
\end{figure}

\begin{figure}
\centering
\includegraphics[width=\linewidth]{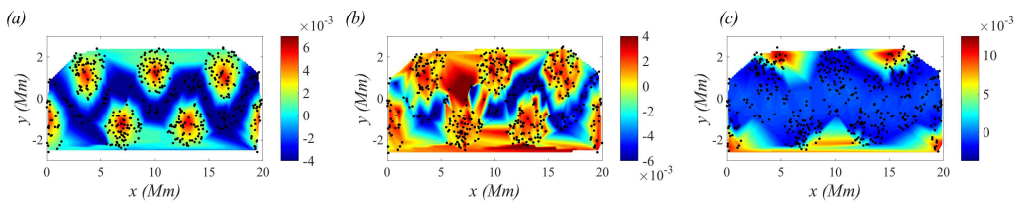}
\caption{CSC contours for the Bickley jet flow overlaid with dots indicating initial positions of 500 drifters for $t=[0$, $6912\times10^3]$ s. (a) ``Real'' trajectories. (b) Simulated trajectories with temporal discretization and stochastic position error, but correct value of $\epsilon(3)=0.3$. (c) Simulated trajectories with temporal discretization, stochastic position error, and incorrect value of $\epsilon(3)=0$.}
\label{fig:bickley3}
\end{figure}

\begin{figure}
\centering
\includegraphics[width=\linewidth]{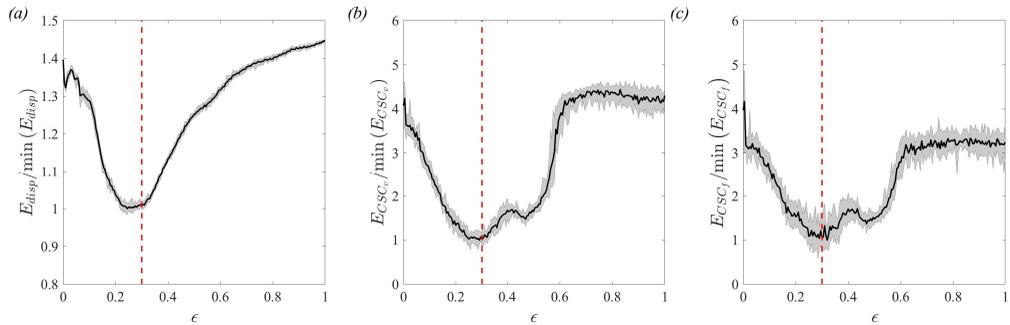}
\caption{Normalized error versus candidate parameter $\epsilon(3)$.  Error metrics are normalized by the lowest error over the range of $\epsilon(3)$ tested.  Solid black line indicates the error averaged over 10 individual sweeps, and shaded region bounds the range of errors seen for an individual sweep.  Red dotted line indicates the correct value for the parameter $\epsilon(3)$. (a) particle displacement error, i.e. eq.~\ref{eqn:disp_error} (b) CSC vector error, i.e. eq.~\ref{eqn:CSC_vector_error} (c) CSC field error, i.e. eq.~\ref{eqn:CSC_field_error}}
\label{fig:bickley4}
\end{figure}

The second test case uses the Bickley jet flow to highlight the effect that temporal discretization and position error, both initially and along the drifter trajectory, have on the process of determining model parameters accurately.  Figure~\ref{fig:bickley1} shows a sample of ten of the 500 total trajectories for the ``real" set of drifters (left); the simulated set with the correct value of $\epsilon(3)=0.3$, with temporal discretization but no position error (center); and the simulated set with the correct value of $\epsilon(3)=0.3$, where both temporal discretization and position error were included (right). 
Temporal discretization and stochastic position error both tend to push drifters near the centers of the gyre cores toward the edge of the gyres, or even out completely into the background flow, although in all three cases highlighted in figure~\ref{fig:bickley1}, the zonal meandering jet and gyre cores remain distinct.  It is clear that the both temporal discretization and stochastic position error will both act to degrade the effectiveness of the error analysis, and if pushed to extremes, will render \textit{any} error metric ineffective at parameter estimation.  The question, in this case, is whether the minimum in the displacement error and/or CSC field error are robust to reasonable levels of these effects.  One potentially useful technique in countering the effects in this study is selective drifter placement.  Here, the set of ``real" drifters were initialized near the center of the coherent structures in the flow, as dictated by the ridges in the FTLE field.  Figure~\ref{fig:bickley2} shows the selected ``real'' drifter initial positions, with the FTLE field in the background.  Practically, it might be challenging to know where to initialize real drifters in an oceanic or atmospheric flow, but this could potentially be accomplished using iterative deployments: the first to determine drifter placement for the second, and the second set of drifters used to tune model parameters.
Figure~\ref{fig:bickley3} shows the CSC field for the real set of particles (left), the simulated set of particles with temporal discretization and stochastic position error with the correct value of $\epsilon(3)=0.3$ (middle), and the set simulated set of particles with an incorrect value of $\epsilon(3)=0$ (right).  For both the real set of trajectories and the simulated set with the correct value of $\epsilon(3)$, the flow is dominated by the meandering jet and the flanking vortices.  Despite the addition of error in trajectory calculation, the CSC field for the simulated sets of trajectories still exhibit distinct coherence.  For the incorrect values of $\epsilon(3)$, the jet has a much wider vertical extent, and only portions of the flanking vortices are seeded, while the extent of the jet for the correct value of  $\epsilon(3)$ is similar to that of the real flow.  
The particle displacement error, CSC vector error, and CSC field error for the full range of tested parameters are shown in figure~\ref{fig:bickley4}.  As with the previous study, each parameter value was tested using ten individual simulations, and the average and range of error for each parameter value is shown in the figure, along with the correct value of $\epsilon(3)=0.3$. The same trends seen in test case 1 are also present here.  The minimum error in the CSC field is very close to the correct value of $\epsilon(3)$, with identified values of $\epsilon(3)$ ranging from 15\% below the real value to 9\% above.  However, the error metric based on particle displacement identifies values of $\epsilon(3)$ ranging from 7\% to 20\% below the real value.  In this case, the CSC vector error comes close to estimating the unknown parameter correctly, in contrast with the study focusing solely on initial particle error.  Additionally, the CSC error metric enhances the difference in error between correct and incorrect parameter values over particle location error, as with the previous study (i.e. compare the vertical axes of the panels in figure~\ref{fig:bickley4}). 

\begin{figure}
\centering
\includegraphics[width=\linewidth]{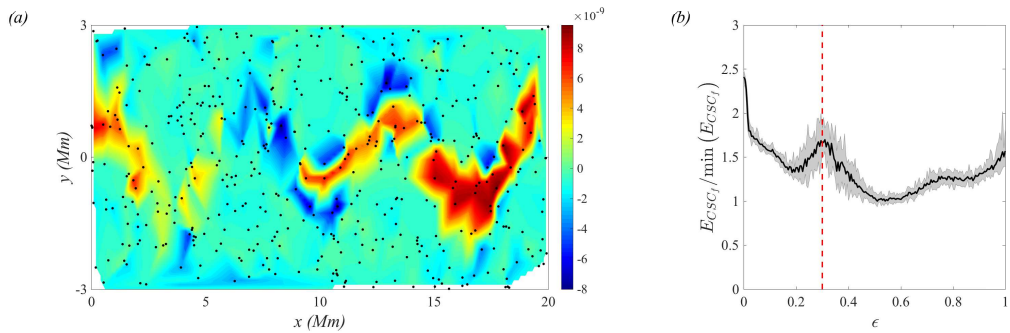}
\caption{Effects of random particle distribution (a) CSC field for simulated trajectories with temporal discretization, initial error, and path error and a correct value of $\epsilon(3)=0.3$. (b) Normalized CSC field error, i.e. eq.~\ref{eqn:CSC_field_error} versus candidate parameter $\epsilon(3)$.  Error metric is normalized by the lowest error over the range of $\epsilon(3)$ tested.  Solid black line indicates the error averaged over 10 individual sweeps, and shaded region bounds the range of errors seen for an individual sweep.  Red dotted line indicates the correct value for the parameter $\epsilon(3)$.}
\label{fig:bickley5}
\end{figure}

It is useful to understand the effects of using a random drifter distribution instead of a set of drifters initialized inside the coherent structures in the flow.  Figure~\ref{fig:bickley5} shows the CSC field for simulated set of trajectories with a correct value of $\epsilon(3)$ (panel a) and the CSC field error for a range of epsilon values, again using a random initial particle distribution, and otherwise identical parameters to the study discussed in this section.  Due to the lack of particles remaining in the vortex cores, the CSC analysis is unable to detect the kinematic dissimilarity between the vortices and the jet, and instead identifies as most dissimilar the drifters in the jet from those directly outside the jet.  Due to this, the CSC field error is unable to correctly identify the model parameter being analyzed.  This highlights the importance of understanding the effects driving real and simulated trajectories apart, and how to mitigate these effects on the CSC field for simulated particles by using appropriate drifter placement.

\begin{figure}
\centering
\includegraphics[width=\linewidth]{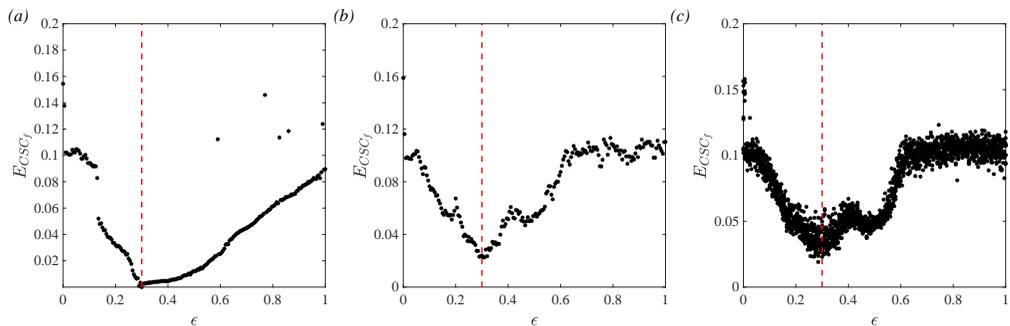}
\caption{CSC field error, i.e. eq.~\ref{eqn:CSC_field_error} versus candidate parameter $\epsilon(3)$.  Error metric is not normalized in this case.  Red dotted line indicates the correct value for the parameter $\epsilon(3)$. (a) trajectories advected with no temporal discretization and no path or initial error (b) trajectories advected with temporal discretization and no path or initial error (c) trajectories advected with both temporal discretization, path error, and initial error, data collected over ten independent parameter sweeps}
\label{fig:bickley6}
\end{figure}

It is also necessary to consider the relative importance of the individual sources of error considered in this analysis.  To this end, figure~\ref{fig:bickley6}(a) shows the CSC field error for simulated trajectories without both temporal discretization and stochastic position error. In this case, the trajectories are deterministic, and the error at the correct parameter value of $\epsilon(3)=0.3$ is identically zero, because the simulated trajectories perfectly match the real set of trajectories.  Figure~\ref{fig:bickley6}(b) shows the CSC field error with temporal discretization but no initial or path error, and figure~\ref{fig:bickley6}(c) includes both temporal discretization and stochastic error.  
When considering only temporal discretization, the trajectories are deterministic given a value of the unknown parameter $\epsilon(3)$.  The CSC field error for temporal discretization only has a global minimum at the correct value of $\epsilon(3)=0.3$, but the error values are generally higher, and the shape of the error curve is different, exhibiting several local minima.  The difference between figures~\ref{fig:bickley6}(b) and (c) shows that the addition of initial and path error to temporal discretization serves to increase the noise in the error curve, which leads to the spread of identified parameter values around the correct value, as discussed previously.  It is important to note that while the parameters chosen for temporal discretization and position uncertainty do lead to some error in the identification of the test parameter, the CSC field error still provides a better metric for identifying unknown parameters than the particle displacement error.

\section{Conclusions}\label{sec:conclusions}
The two test cases examined in this study identify and characterize the influence of initialization errors, temporal discretization, and stochastic path error  on the process of Lagrangian data assimilation for model parameter estimation.  By considering a field-based definition of the underlying flow structure using coherent structure coloring, model parameters can be more accurately and robustly determined than by considering particle displacement errors alone.  This study highlights the value of coherent structure identification, and the CSC algorithm in particular, in model parameter estimation.  This method can potentially be extended from the determination of model parameters in analytical flows to the more complex problem of Lagrangian data assimilation into large-scale oceanic and atmospheric models.  The CSC-based method could eliminate the need for simulation ensembles in order to accurately estimate model parameters, thereby advancing the state of the art in numerical flow prediction.

\begin{acknowledgments}
This work was supported by the U.S National Science Foundation and by the Department of Defense (DoD) through the National Defense Science $\&$ Engineering Graduate Fellowship (NDSEG) Program.
\end{acknowledgments}

\bibliographystyle{jfm}
\bibliography{Schlueter}

\end{document}